\documentclass[%
 reprint,
superscriptaddress,
 amsmath,amssymb,
 aps,
floatfix,
]{revtex4-2}

\usepackage{float}
\usepackage{babel}
\usepackage{xcolor}
\usepackage{graphicx}
\usepackage{dcolumn}
\usepackage{bm}
\usepackage{IEEEtrantools}

\begin{document}

\preprint{APS/123-QED}

\title{Reentrant Rigidity Percolation in\\Structurally Correlated Filamentous Networks}

\author{Jonathan Michel}
\email{Corresponding author: jamsps@rit.edu}
 \affiliation{School of Physics and Astronomy, Rochester Institute of Technology}
\author{Gabriel von Kessel}
 \affiliation{School of Physics and Astronomy, Rochester Institute of Technology}

\author{Thomas Wyse Jackson}
\affiliation{School of Physics, Cornell University}

\author{Lawrence J. Bonassar}
\affiliation{Meinig School of Biomedical Engineering, Cornell University}
\affiliation{Sibley School of Mechanical and Aerospace Engineering, Cornell University}

\author{Itai Cohen}
\affiliation{School of Physics, Cornell University}%
\author{Moumita Das}
\affiliation{%
School of Physics and Astronomy, Rochester Institute of Technology}%

\date{\today}

\begin{abstract}
Many biological tissues feature a heterogeneous network of fibers whose tensile and
bending rigidity contribute substantially to these tissues' elastic properties.
Rigidity percolation has emerged as a important paradigm for relating these
filamentous tissues' mechanics to the concentrations of their constituents. Past studies
have generally considered tuning of networks by spatially homogeneous variation in
concentration, while ignoring structural correlation. We here introduce a model 
in which dilute fiber networks are built in a correlated manner that produces
alternating sparse and dense regions. Our simulations indicate that
structural correlation consistently allows tissues to attain rigidity with
less material. We further find that the percolation threshold varies 
non-monotonically with the degree of correlation, such that it decreases with
moderate correlation and once more increases for high correlation.
We explain the eventual reentrance in the dependence of the rigidity
percolation threshold on correlation as the consequence of large, stiff clusters
that are too poorly coupled to transmit forces across the network. Our study
offers deeper understanding of how spatial heterogeneity may enable tissues to
robustly adapt to different mechanical contexts.
\end{abstract}

\maketitle


\section{\label{sec:intro} Introduction}
Networks and network-like structures are ubiquitous in biological cells and tissues, and provide the basis for their mechanical properties and functions. Biopolymer networks are largely responsible for the mechanical response of the cytoskeleton of cells \cite{Pollard, Gardel1, Gardel2, Murrell, Lee1, Lee2} and the extracellular matrix of tissues \cite{Munster, silverberg_structure-function_2014, KAJansen, Burla1, Wyse_Jackson}; more recently, rigidly percolating connected networks of cells have been shown to account for the viscoelasticity of developing embryos \cite{Heisenberg}. These networks are generally highly disordered and spatially inhomogeneous as a result of how they are assembled and disassembled. For example, cytoskeletal networks are highly dynamic and have a complex and heterogeneous spatial organization allowing for context-dependent cell remodeling and response \cite{Pollard}.  As a second example, the collagen II scaffold in articular cartilage is densest in the vicinity of chondrocytes, cells which secrete extracellular matrix material to construct and sustain collagen networks \cite{DiDomenico}.  In fact, previous work has established spatial heterogeneity as a crucial consideration in developing faithful cartilage replacements and scaffolds for tissue regeneration \cite{Rhee, Sridhar}.

In the last two decades, there has been extensive study of disordered biopolymer networks through in-vitro experiments and simulations, which have provided a wealth of information about these networks' responses to mechanical stimuli \cite{Head, Wilhelm, Heussinger, Das1, Huisman, Broedersz1, Das2, Picu, Broedersz2}. To date, however, almost all computational studies of biopolymer networks have focused on spatially homogeneous systems and ignored the presence of structural correlations, which can have significant consequences for the collective properties of the network. 
Here, we address this gap and present a novel investigation of the percolation of rigidity in structurally correlated fiber networks, which are found in many cells and tissues, using a lattice-based framework. 


Lattice-based fiber networks are a prominent paradigm for modeling biopolymer scaffolds \cite{Feng, Das1, Broedersz1, Das2, Broedersz2}. These networks are constructed by laying down infinitely long fibers in a regular pattern, such that whenever two fibers cross, there is a crosslink which allows free rotation of the fibers but does not allow them to slide. The fibers can stretch and bend, but pay energy penalties for these deformations.  Each fiber can further be thought of as a collinear series of connected bonds, such that random removal of
bonds yields a broad distribution of fiber lengths. 

The mechanical response of such a disordered fiber network can be mapped to the fraction of bonds present. Starting with a network in which all the bonds are present, one can progressively decrease network rigidity by removing bonds. Once the network reaches a certain threshold of bond occupation, its elastic moduli undergo a dramatic, many-decade decrease, dropping to negligibly small values. This mechanical phase transition is known as rigidity percolation\cite{Feng,Das1}. Dilute fiber networks have shown great promise as micromechanical models for in-vitro cytoskeletal networks \cite{Head, Wilhelm, Heussinger, Das1}, and more recently for extracellular matrix networks in tissues ~\cite{Storm, silverberg_structure-function_2014, Wyse_Jackson}.

Previous computational studies of fiber networks have examined spatially homogeneous disordered networks, in which bonds are excluded or retained purely at random, and have not considered the possibility of correlations in the inclusion of bonds \cite{Head, Wilhelm, Das1, Huisman, Broedersz1}. In this paper we introduce a model in which bonds are added in a structurally correlated manner to account for the pronounced heterogeneity in the distribution of material observed in cells and tissues. In this model, the
likelihood of adding a bond is contingent upon the number of adjacent bonds
already present. This protocol gives rise to networks in which already dense regions become further enriched with material, while sparse regions remain comparatively dilute. 

\section{\label{sec:model} Model}

\subsection{Network Construction}

 As we are interested in biopolymer networks in which vertices correspond to crossings of adjacent filaments, we choose as our model network the kagome lattice \cite{silverberg_structure-function_2014}, with a maximum coordination number of 4.  We adjust the elastic moduli of networks by randomly retaining a subset of the bonds, such that some portion, $p$, is included. In the absence of structural correlation, $p$ corresponds the probability that each
bond is retained, such that there is an independently and identically distributed probability of keeping each bond. We instead use the
term bond portion to reflect the fact that, for correlated dilute
networks, once the network is seeded with an initial set of bonds,
other bonds do not have an identical likelihood of inclusion.

\begin{figure*}
    \centering
    \includegraphics[width = 17.2cm]{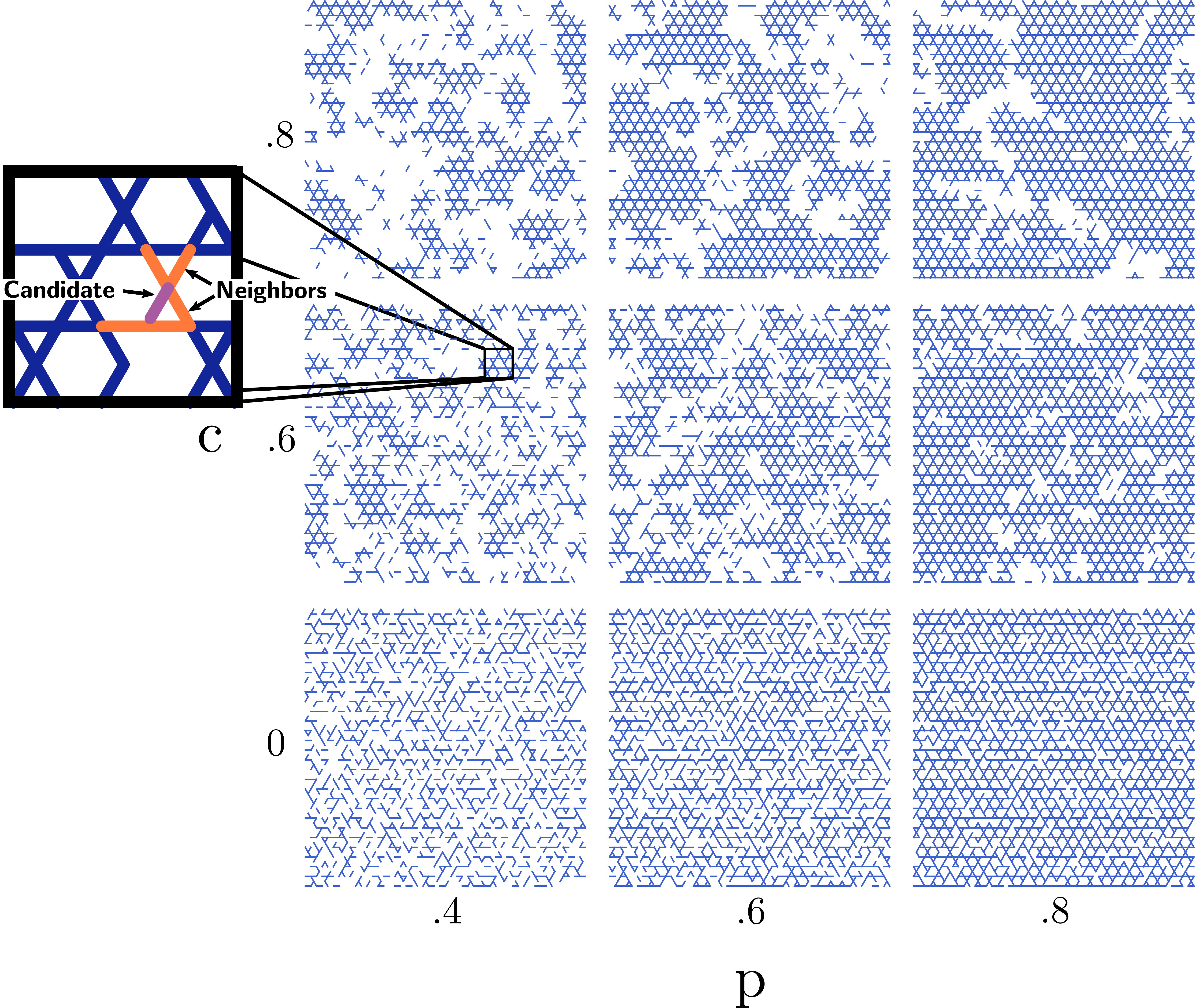}
    \caption{
    \label{corr_comp}
    Sample networks are shown in which correlation
    strength is varied from 0 to .8, and the bond portion is
    varied from .4 to .8. As the correlation strength is 
    increased, networks transition from arrangements of 
    homogeneously dispersed bonds to networks with dense clusters
    interspersed with sparse voids. While individual clusters may
    be rigid, they may be tenuously connected to neigboring 
    clusters, such that, beyond some critical correlation 
    strength, networks will become less effective in transmitting
    force over macroscopic distances. In the magnified inset at the left of the
    figure, we illustrate the correlated construction process. A candidate,marked in
    magenta, is considered for inclusion, and has five neighbors, marked in orange.}
\end{figure*}

We employ an iterative process, introduced in ~\cite{zhang_corr}, in which, at each step, a candidate bond is chosen at random from those bonds that have not yet been included in the network. We then count the number of bonds adjacent to the candidate bond, $n_a$ that have already been retained, where adjacency is defined by the condition that two bonds share a common vertex. For the kagome lattice, the maximum possible number of adjacent bonds, $n_{a, max}$, is 6. Given a correlation strength, c, where $0 \leq c < 1$, the candidate bond is added with a probability

\begin{equation}
    P = (1 - c)^{n_{a, max} - n_a},
\end{equation}
where $c = 0$ corresponds a purely random dilution, and $c \approx 1$ corresponds to  maximally correlated 
dilution, in which a bond is rejected unless all adjacent bonds have been retained. This process is repeated until the desired portion of bonds has been retained. In figure
\ref{corr_comp}, we show representative samples of networks with varying degrees of dilution and correlation, including no correlation. 

\subsection{Mechanical Model}

The bonds of the network resist stretch and compression with a Hookean spring stiffness, $\alpha$, and
resist bending with a bending rigidity, $\kappa$. Bending resistance is implemented by regarding
adjacent, collinear bonds as consecutive segments of a fiber. If the bond connecting vertices $i$ and $j$ is
collinear with and adjacent to the bond joining vertices $j$ and $k$, then we penalize a change in the
angle $\angle ijk$ by an amount proportional to the square of the angular deflection. 


We focus on the mechanical response in the linear response regime, so that 
the deformation energy consists of terms quadratic in the strain.  Following 
~\cite{silverberg_structure-function_2014}, we truncate the deformation energy to
leading order in the displacement of vertices from the reference configuration of
the network, and model the energy as

\begin{align}
    \label{eq:efunc}
    E_{strain} &= \frac{\alpha}{2} \sum_{<ij>} p_{ij} (\mathbf{u}_{ij} \cdot \hat{\mathbf{r}}_{ij})^2 \nonumber \\
    &+ \frac{\kappa}{2}\sum_{<ijk>} p_{ij} p_{jk} [(\mathbf{u}_{ji}+\mathbf{u}_{jk}) \times \hat{\mathbf{r}}_{ji}]^2 \nonumber \\
\end{align}
Here, $<ij>$ denotes a sum over pairs of vertices sharing a bond, $<ijk>$ denotes a sum over vertices
of adjacent, colinear bonds, and $p_{ij}$ is defined to be 1 if the bond between bonds $i$ and $j$ is
retained, and 0 otherwise. Further, $\mathbf{u}_{ij}$ denotes the difference between the displacement
vectors for vertices $i$ and $j$, and $\hat{\mathbf{r}}_{ij}$ denotes the direction vector of the
bond between vertices $i$ and $j$ in the reference state.

\subsection{Structural Relaxation Procedure}

We simulate the shear mechanics of our model networks by imposing a series of small, simple shear displacements of the vertices at the top. The 
displacements of the vertices at the bottom of the network are constrained to be zero, while along the sides we impose periodic boundary
conditions. Given these constraints, we minimize the energy given in \eqref{eq:efunc} for several different
shear strains, $\epsilon_s$, and compute the shear modulus at a target shear strain, $\epsilon_s^*$ by
numerically estimating the second derivative of the strain energy with respect to shear strain:

\begin{equation}
    G = \frac{\partial^2 E_{strain}}{\partial \epsilon_s^2} \Big|_{\epsilon_s^*}.
\end{equation}
Full details of this calculation may be found in Appendix \ref{relax_appendix}.

\section{\label{sec:res} Results and Discussion}

\subsection{Shear Mechanics}

We first examined how the rigidity percolation threshold can be tuned by varying 
the degree of correlation. We considered nine distinct values of $c$, ranging from
0 to $0.8$, in steps of $.1$, and 61 distinct values of
$p$, ranging from $0.4$ to $1$, in steps of $0.05$. We 
carried out structural relaxation for $10$ realizations
for each combination of $c$ and $p$, and identified $G$ for
each combination as the geometric mean of the $10$ values. The shear
modulus $G$, normalized by its universal maximum, $G_0$, is shown vs.
$p$ for several values of $c$ in Figure 2\textbf{a}, 
accompanied by a full phase diagram in Fig. 2\textbf{b}.

\begin{figure}
    \centering
    \includegraphics[width=8.6cm]{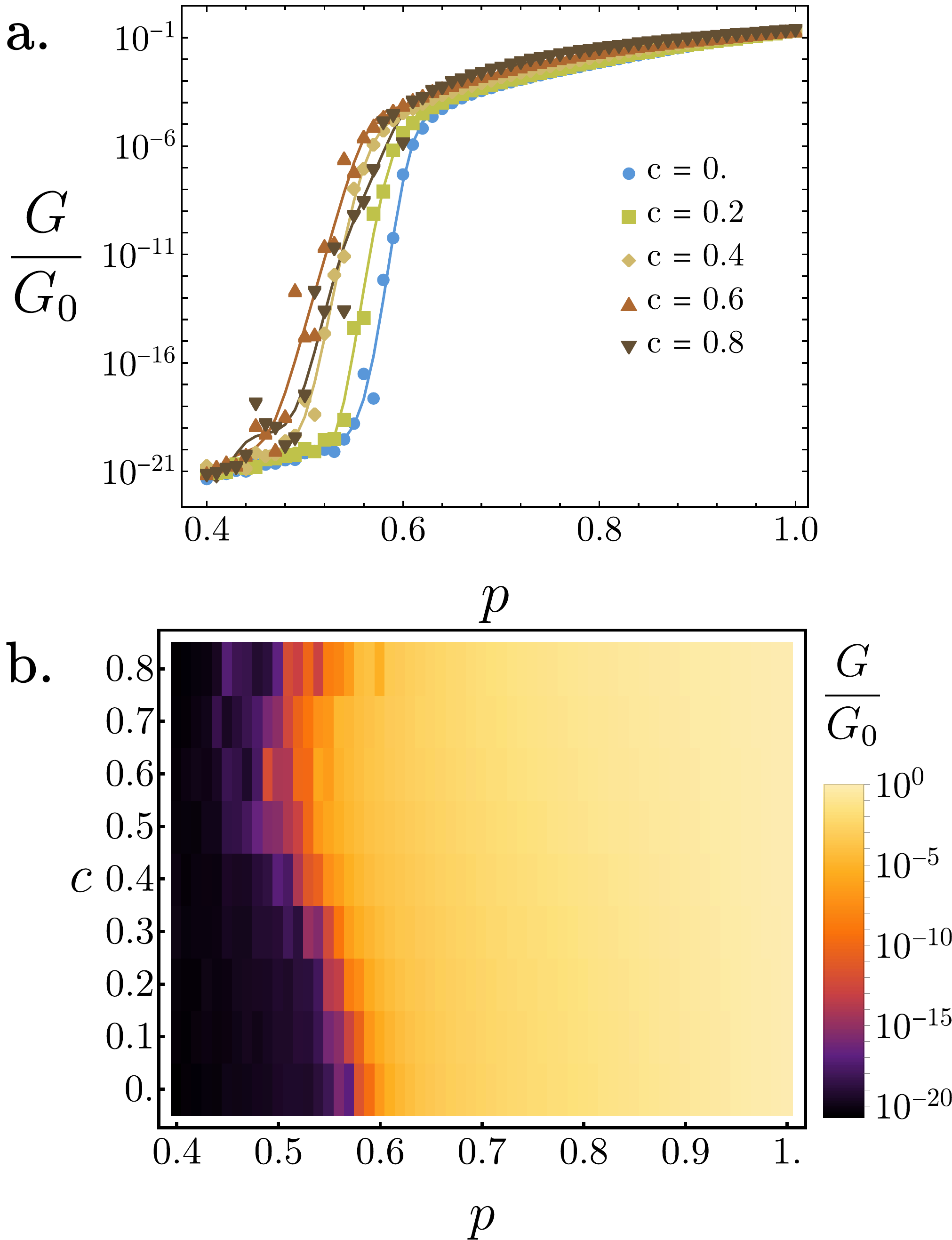}
    \caption{
    \label{fig:gmod_results}
    In panel \textbf{a}, we show the scaling of shear modulus with bond portion
    for several structural correlation strengths. While the
    dependence of the shear modulus on the bond portion is
    qualitatively similar in each case, the point of rigidity
    percolation shifts initially to the left, then back to
    the right with increasing structural correlation. In panel \textbf{b}, we show the dependence
    of G on $c$ and $p$ for the full range of parameter space considered. The reentrance of the dependence of $p_c$ on
    structural correlation strength is clearly discernible
    in a contour of marginal stiffness on the left side of
    the heat map. We attribute this reentrance to two competing
    effects: the need for rigid clusters, and
    the need for strong coupling between adjacent clusters.}
\end{figure}

Figures 2\textbf{a} and 2\textbf{b} provide the first indications of an intriguing variation in the rigidity percolation threshold with the degree of correlation $c$ of the disordered network.
For each value of $c$, we find a qualitatively similar scaling of the shear modulus with the bond fraction. Interestingly, however, the rigidity percolation threshold i.e., the critical bond fraction, $p_c$, 
at which the shear modulus first differs appreciably from zero, shifts markedly and non-monotonically as $c$ is varied: While introducing a moderate correlation strength
initially diminishes $p_c$, this effect saturates at about $c = .6$, and $p_c$ increases for still larger values of $c$.

To quantitatively identify the rigidity percolation threshold for each value of $c$, we considered pairs of bond fraction and
shear modulus for which the shear modulus ranged from $10^{-9}$ to $\sim 10^{-12}$. This ensured that the shear modulus was greater than machine or 
algorithmic error, but still small in comparison with its maximal value, $G_0$, at $p = 1$. We used the method of least squares to fit each set of bond fraction-shear 
modulus pairs to a power law of the form

\begin{equation}
    \label{eq:power_law_fit}
    G = k \left(p - p_c\right)^{\beta}.
\end{equation}
For each value of $c$, we found a good fit to equation ~\ref{eq:power_law_fit} over over at least
seven decades of dynamic range in the shear modulus. In each case, the correlation coefficient, $R^2$, between $\log_{10}(G)$ and $\log_{10}[k(p - p_c)^{\beta}]$ is $\geq .96$. Further details
are provided in Appendix \ref{smod_appendix}.

As shown in Fig. ~\ref{fig:power_law_fits}\textbf{a}, in which $p_c$ is plotted vs. $c$, our power law fits affirm the trend in $p_c$ previously identified by inspection in Fig 2. 
Surprisingly, the scaling exponent, $\beta$, on the other hand, exhibits the opposite trend, increasing with $c$ until about $c = .6$, and 
decreasing thereafter. This unexpected scaling of $\beta$ can be explained as an earlier onset
of rigidity percolation being associated with a more abrupt rise in the shear modulus at the 
point of percolation. 

As shown in figure ~\ref{fig:power_law_fits}\textbf{c}, we find $\beta$ to decrease linearly with $p_c$ ($R^2 = .92$). We attribute the decrease in $\beta$ with
$p_c$ to two competing factors determining percolation: the presence of
large, rigid clusters, and sound mechanical coupling of adjacent clusters.
As the correlation strength is increased beyond its optimum value, the
network segregates into large, dense regions that are too poorly connected
to enable optimal transmission of stress. The optimal correlation strength
of $.6$ strikes an ideal balance, enabling the greatest gain in stiffness
per unit material. To substantiate this conjecture, we turn to detailed
analysis of network displacement fields.


\begin{figure*}
    \centering
    \includegraphics[width=17.2cm]{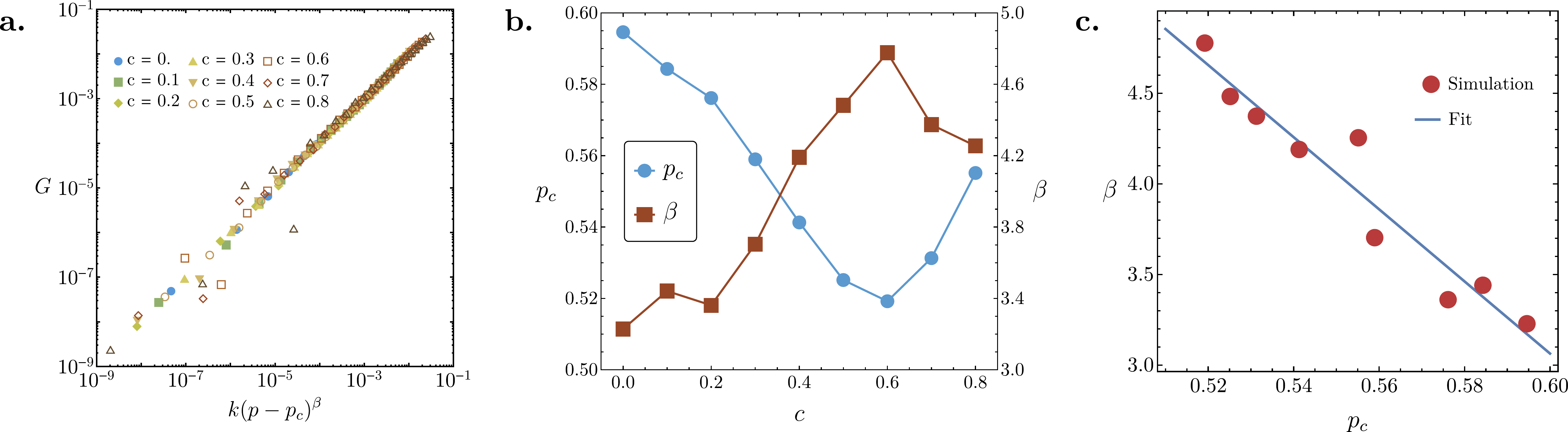}
    \caption{\textbf{Scaling behavior near the rigidity
    threshold: }In \textbf{a}, we show the results of fitting shear modulus - bond
    portion pairs to Eq. ~\ref{eq:power_law_fit}. In each case, we find a sound fit
    spanning 6 to 7 decades in the shear modulus. In \textbf{b}, we show fitting
    results for the critical bond portion, $p_c$, and the scaling exponent, $\beta$, of
    the shear modulus with the excess bond portion. The scaling exponent exhibits the
    opposite trend from the critical bond portion, indicating that a low 
    percolation threshold is accompanied by a more abrupt increase in the shear
    modulus. In \textbf{c}, we demonstrate that the critical bond portion
    is a reliable predictor of the critical exponent relating growth in the 
    shear modulus to the excess bond portion. We find that the value c = .6
    yields an optimal trade-off between a need for large, rigid clusters, and a
    need for sound mechanical coupling of adjacent clusters to enable
    coordinated, system-spanning force propagation. For excessive correlation,
    dense clusters amount to stiff inclusions in an otherwise under-coordinated network.}
    \label{fig:power_law_fits}
\end{figure*}

\subsection{Analysis of Non-affine Deformation}

To gain further insight into the micromechanical mechanisms
underlying the reentrance in the rigidity percolation
threshold, we quantified the degree of non-affinity in
each network. Non-affinity quantifies the departure of
a displacement field of a strained material from the
displacement expected in a simple, homogeneous elastic
continuum. In an affinely deforming network subjected
to simple shear, a vertex with an initial location
$\vec{r}_0 = (x_0, y_0)$ will be mapped to the final 
location $\vec{r}^{\hspace{2pt} \prime}_A = \left(x_0 + \epsilon_s y_0, y_0 \right)$, where $\epsilon_s$ is the
shear strain. The non-affine displacement field, 
$\vec{u}_{NA}$ is defined as $\vec{r}^{\hspace{2pt} \prime} - \vec{r}^{\hspace{2pt} \prime}_{NA}$, where
$\vec{r}^{\hspace{2pt} \prime}$ is the true displacement
field. The non-affine parameter, $\Gamma$, is then defined
as

\begin{equation}
    \label{eq:nadef}
    \Gamma = \frac{1}{N \hspace{1pt} \epsilon_s^2 \hspace{1pt} l_0^2} \sum_{i = 1}^N \left| \vec{u}_{NA} \right|^2,
\end{equation}
where $l_0$ is the length of an undeformed bond, and $N$ is
the number of vertices in the network \cite{DiDonna}.
Non-affinity is a well established means of characterizing
the difference in behavior of a purely entropic
rubber and a gel of semiflexible polymers \cite{Broedersz2}.

\begin{figure*}
    \centering
    \includegraphics[width=17.2cm]{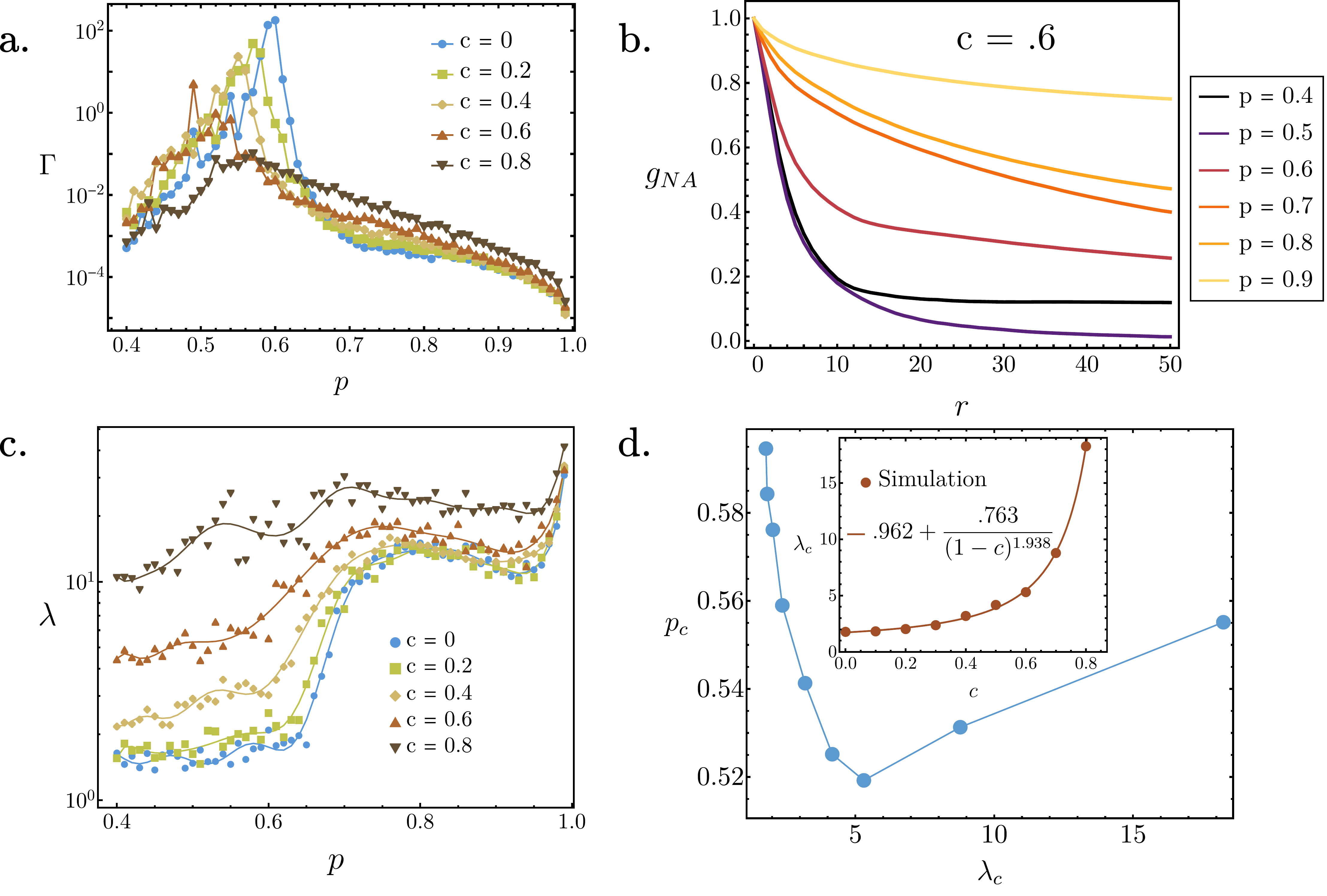}
    \caption{\textbf{a}. We show the non-affine parameter, as defined in
    Eq.~\ref{eq:nadef}, vs bond portion for a number of correlation
    strengths. In each case, the non-affine parameter exhibits a 
    pronounced peak at the rigidity percolation transition. Larger correlation
    strengths correspond to non-affine re-arrangements occurring over a wider range of 
    bond-fractions and are reflected in the peaks becoming progressively broader with increasing $c$. \textbf{b} We display azimuthally averaged non-affine correlations $g_{NA}(r)$ as a function of distance $r$ between vertices. In each case, we find the decay to be exponential, allowing us to extract an emergent mechanical length scale, $\lambda$. This length scale monotonically increases with bond portion $p$, suggesting longer range correlations for networks with more fiber content. The non-affine correlations in the large $r$ limit decrease with increasing bond portion below the rigidity percolation threshold; however, above this threshold they increase with increasing bond portion. While we choose a correlation strength of .6 here, networks with differing correlation strength exhibited qualitatively similar behavior. 
    In \textbf{c}, we show emergent mechanical length scales $\lambda$ for different combinations of $p$ and $c$. At low bond portions, the length scale $\lambda$ steadily increases with correlation strength for a given $p$, whereas at high bond portions, values of $\lambda$ for different values of $c$ converge and approach the system size. In \textbf{d}, we identify a critical mechanical length scale,
    $\lambda_c$, as the decay length of non-affine correlations at the onset of rigidity
    percolation, which exhibits a power law divergence as the structural correlation strength nears 1 (see inset). We show that the critical bond portion $p_c$ varies non-monotonically with $\lambda_c$ in a manner reminiscent of the scaling of $p_c$ with with structural correlation strength $c$ in Fig. 3(b).}
    \label{fig:na_scaling}
\end{figure*}

We computed $\Gamma$ for all ten network realizations for
each combination of $p$ and $c$, and calculated the final 
non-affine parameter as the arithmetic mean over all 
realizations. As reported in previous studies, we find
a pronounced peak in the non-affine parameter near the
rigidity percolation threshold, as shown in Figure
\ref{fig:na_scaling}\textbf{a}; however, for increasing
structural correlation, this peak becomes progressively 
lower and broader. We posit that this peak broadening
is associated with the formation of local rigid regions,
joined by weakly connected interstitial regions, such
that stress is distributed in a non-uniform manner over a
larger range of bond fractions for networks with high
structural correlation. We investigate this idea further by
considering the dispersion in the distribution of bond
strains, and the spatial correlations in the non-affine
displacement field.

We first consider a radial non-affine correlation function,
$g(r)$. We take the inner product of non-affine displacements for
all pairs of points within some cutoff distance, $r_{cut}$, of one 
another, and bin displacement vectors between pairs of
points into annular sectors of thickness $\Delta r$. We then
define $g(r)$ as

\begin{equation}
    \label{eq:rad_na_corr}
    g(r) = \frac{1}{\left< \left|\vec{u}_{NA} \right|^2 \right>} \times
    \Big< \vec{u}_{NA}(\vec{r}_i) \cdot \vec{u}_{NA}(\vec{r}_j)\Big>_{r \leq |\vec{r}_j - \vec{r_i}| < r + \Delta r},
\end{equation}
where the first average runs over all points, and the second
average runs over all distinct pairs of points $i, j$ such
that the positions of vertices $i$ and $j$ in the undeformed
lattice are separated by a distance in the range
$[r, r + \Delta r)$. This normalizes $g(r)$ to be equal to
1 when $r = 0$.

For each combination of $p$ and $c$, we find $g(r)$ to be
well fit ($R^2 \geq .98$) by the form

\begin{equation}
    \label{eq:g_of_r_fit}
    g(r) = 1 + a\left(e^{-r / \lambda} - 1\right).
\end{equation}
We show a representative set of curves in Fig. 
~\ref{fig:na_scaling}\textbf{c},
for $c = .6$ and varying bond fractions. Initially, the
floor of $g(r)$ decreases, reaching its lowest point near
the rigidity percolation threshold, then steadily increases
for larger values of $p$. We attribute the early decrease
in the floor to incipient rigid clusters that deform
differently from the surrounding soft regions, such that
there is no coordinated, long-range force transmission. The
subsequent rise in the minimum of $g(r)$ is associated with
the emergence of system-wide spanning force chains beyond
the rigidity percolation threshold.

From the decay distance, $\lambda$, in equation ~\eqref{eq:g_of_r_fit}, we infer an
effective mechanical length scale. Different correlation
strengths yield qualitatively similar dependence of $\lambda$
upon bond portion, but $\lambda$ at a fixed bond portion
steadily increases with increasing correlation. This further
affirms the idea that the size of a coherently deforming
cluster becomes progressively larger with growing correlation.
Results are shown in Figure ~\ref{fig:na_scaling}, with trend
lines computed from cubic basis splines.

We finally seek to account for the reentrant scaling of $p_c$ with
structural correlation strength by identifying the critical mechanical
length scale, $\lambda_c$, at the onset of rigidity percolation for
each structural correlation strength, $c$. We estimate $\lambda_c$ by interpolation
using the previously mentioned cubic basis splines.  We find that $p_c$ varies
non-monotonically with $\lambda_c$, with an initial decrease until $\lambda_c$ exceeds
about 5 bond lengths, after which $p_c$ once more increases. The optimal value
of $\lambda_c$ is that obtained for a structural correlation strength of .6, in
concert with our previous findings. We further observe that $\lambda_c$ appears
to diverge according to a power law as $c$ approaches 1. In the limit $c = 1$,
either all bonds can be present or no bonds can be present, so that the only
percolating network would be a fully connected network, in which vertex displacements
are correlated over arbitrarily large distances. We thus find that, while small,
rigid islands must nucleate to enable the most efficient percolation, excessively
large rigid clusters leave too little material elsewhere to enable the formation
of system-wide force chains.


\section{Conclusion}

We have introduced and investigated a model of rigidity percolation in spatially correlated
networks. Our study of the scaling of the shear modulus near percolation, coupled with
our analysis of networks' strain fields, offers a straightforward physical
picture accounting for the reentrant scaling of $p_c$ with $c$. While the length
scale over which a network's displacement field is well coordinated grows
monotonically with correlation strength, eventually neighboring rigid clusters
become poorly coupled. Weak tethers between dense islands of bonds lead to
strain being highly concentrated, rather than the load being distributed evenly
throughout the network. 


This work broadens the already successful rigidity percolation framework to better account for the mechanical response of structurally correlated, heterogeneous networks found 
in cells and tissues. We anticipate this work will usher in further studies exploring the role of anisotropy ~\cite{Zhang_Aniso} observed in many extracelluler matrices. 
Our findings indicate that, rather using just an averaged, system-wide characterization of network topology, local 
spatial patterns should be considered to fully understand tissues' responses to
applied stress.

\appendix{}

\section{Finding Mechanical Ground States of Elastic Networks}
\label{relax_appendix}

We seek a zero-force configuration of the network, subject
to the constraint that the the nodes along the bottom of the
network are fixed, and the nodes along the top of the 
network do not translate in the y direction, and are 
displaced by a uniform amount to the right. Periodic
boundary conditions are imposed at the left and right
boundaries of the network. We note that,
due to the quadratic energy given by Eq. ~\ref{eq:efunc},
the restoring force, $\vec{F}$, resulting from a
displacement field $\vec{u}$ may be computed as

\begin{equation}
    \vec{F} = -\mathbf{K}\vec{u},
\end{equation}
where, for a network with $N$ vertices, $\mathbf{K}$ is
a $2N \times 2N$ matrix, and $\vec{u}$ is a 
$2N \times 1$-dimensional column vector with the 
displacement field components of node $i$ given in indices
$2i - 1$ and $2i$ of $\vec{u}$. For a total energy $E$,
the matrix element $K_{\alpha \beta}$ is given by

\begin{equation}
    K_{\alpha \beta} = \frac{\partial^2E}{\partial u_{\alpha} \partial u_{\beta}},
\end{equation}
where $1 \leq \alpha, \beta \leq 2N$.

We then partition indices of the displacement field on the interval $[1, 2N]$ into two subsets: the set
$\mathcal{R}$ of indices corresponding to relaxed 
coordinates, and the set $\mathcal{B}$ of indices 
corresponding to constrained coordinates on the boundary.
Let $\vec{u}_R$ be an $R \times 1$ column vector containing
just those coordinates permitted to relax, where 
$R = |\mathcal{R}|$ is the number of relaxed coordinates.
We further define a projection operator from the full 
$2N$-dimensional displacement field $\vec{u}$ to 
$\vec{u}_R$, denoted by $\mathbf{P}_{N \rightarrow R}$,
and a projection operator $\mathbf{P}_{R \rightarrow N}$ 
from $\vec{u}_R$ back to $\mathbb{R}^{2N}$. The product
$\mathbf{P}_{R \rightarrow N} \mathbf{P}_{N \rightarrow R}$
yields a $2N \times 2N$ linear operator satisfying

\begin{equation}
    \left(\mathbf{P}_{R \rightarrow N} \mathbf{P}_{N \rightarrow R}\right)_{\alpha \beta} =
    \left\{\begin{array}{rl}
        1, & \alpha = \beta \hspace{2pt} \text{and} \hspace{2pt} \alpha \in \mathcal{R}\\
        0, & \hspace{2pt} \text{otherwise}
    \end{array}\right..
\end{equation}
Finally, we define a $2N \times 2N$ operator, 
$\mathbf{I}_B$, to select just those elements of $\vec{u}$
corresponding to boundary nodes' displacements:

\begin{equation}
    I_{\alpha \beta} =
    \left\{\begin{array}{rl}
        1, & \alpha = \beta \hspace{2pt} \text{and} \hspace{2pt} \alpha \in \mathcal{B}\\
        0, & \hspace{2pt} \text{otherwise}
    \end{array}\right..
\end{equation}

With the foregoing definitions in hand, we now return to
the physical situation. The net force on the relaxed nodes 
due to interactions amongst relaxed nodes must
be the opposite of the net force on relaxed nodes due to
their interaction with boundary nodes. In terms of the
previously defined quantities, this implies:

\begin{equation}
    \mathbf{P}_{N \rightarrow R} \hspace{1pt} \mathbf{K} \hspace{1pt} \mathbf{P}_{R \rightarrow N} \vec{u}_R = -\mathbf{P}_{N \rightarrow R} \hspace{1pt} \mathbf{K} \hspace{1pt} \mathbf{I}_B \vec{u}.
\end{equation}

We choose as our starting guess the affine displacement
field, $\vec{u}_{A}$, and decompose $\vec{u}$ as the sum
of the affine field and the non-affine field, 
$\vec{u}_{NA}$. Defining $\mathbf{K}_R \equiv  \mathbf{P}_{N \rightarrow R} \hspace{1pt} \mathbf{K} \hspace{1pt} \mathbf{P}_{R \rightarrow N}$, we solve for the
non-affine component of $\vec{u}_R$, $\vec{u}_{NA, R}$ as

\begin{multline}
    \vec{u}_{NA, R} = -\mathbf{K}_R^+ \left( \mathbf{P}_{N \rightarrow R} \hspace{1pt} \mathbf{K} \hspace{1pt} \mathbf{I}_B \vec{u}_A \hspace{1pt} + \right. \\ \left. \mathbf{P}_{N \rightarrow R} \hspace{1pt} \mathbf{K} \hspace{1pt} \mathbf{P}_{R \rightarrow N} \vec{u}_{A, R}\right).
\end{multline}
$\mathbf{K}_R^+$ denotes the Moore-Penrose inverse
of $\mathbf{K}_R$ \cite{Moore,Bjerhammar,Penrose}, which
we compute using the package SuiteSparseQR \cite{Davis}.
We finally solve for the overall displacement field as
\begin{equation}
    \vec{u} = \mathbf{P}_{R \rightarrow N} \vec{u}_{NA, R} + \vec{u}_{NA},
\end{equation}
and the residual strain energy as
\begin{equation}
    E = \frac{1}{2} \vec{u}^T \mathbf{K} \vec{u}.
\end{equation}

\section{Computing Shear Moduli and the Critical Bond Portion}
\label{smod_appendix}

We perform shear simulations of ten realizations for each
combination of bond portion and correlation strength. In
each case, we apply five small, evenly spaced shear strains 
$\epsilon_i, 1 \leq i \leq 5$ to the top of the network. Let
the third strain be $\epsilon^{\star}$, and let the spacing
between successive displacements $\delta$. We then 
approximate the differential shear modulus as

\begin{IEEEeqnarray}{rcl}
    G(\epsilon^{\star}) &=& \frac{\partial^2 E}{\partial \epsilon^2}\Bigg|_{\epsilon = \epsilon^{\star}}\\
    &\approx& \frac{1}{12 \delta^2} \Bigg\{16 \left[G(\epsilon^{\star} + \delta) + G(\epsilon^{\star} - \delta)\right]\nonumber\\
    &&- G(\epsilon^{\star} + 2\delta) - G(\epsilon^{\star} - 2\delta) - 30 G(\epsilon^{\star})\Bigg\}\nonumber\\
    &&+ \mathcal{O}(\delta^4)
\end{IEEEeqnarray}

Shear moduli in the vicinity of the rigidity percolation
threshold, $p_c$ are liable to vary by several orders of
magnitude. To cope with these fluctations, we report the
median shear modulus of all ten realizations. Once we have
obtained shear moduli for all bond portions for a given
correlation strength, we estimate $p_c$ by fitting
$G(p)$ to a power law of the form

\begin{equation}
    \label{eq:supp_power_law}
    G = k (p - p_c)^{\alpha}.
\end{equation}
We neglect values of the shear modulus less than $10^{-9}$,
in units of the bond stretching modulus, to avoid numerical
noise. We also neglect the final 20 points in each series,
corresponding to bond portions greater than $.8$, as
the shear modulus depends upon the bond portion in a
qualitatively different manner as it reaches its upper
plateau. For each correlation strength, we obtain a sound
fit of the series of $(G, p)$ pairs  to Eq. ~\ref{eq:supp_power_law} by the method of least squares. In each case, the correlation coefficient $R^2$ for $\log_{10}(G)$ vs. $\log_{10}\left[k (p - p_c)^{\alpha}\right]$ is $\geq .96$.

\section{Quantifying Anisotropy in Non-affine Correlations}
\begin{figure*}
    \centering
    \includegraphics[width=17.2cm]{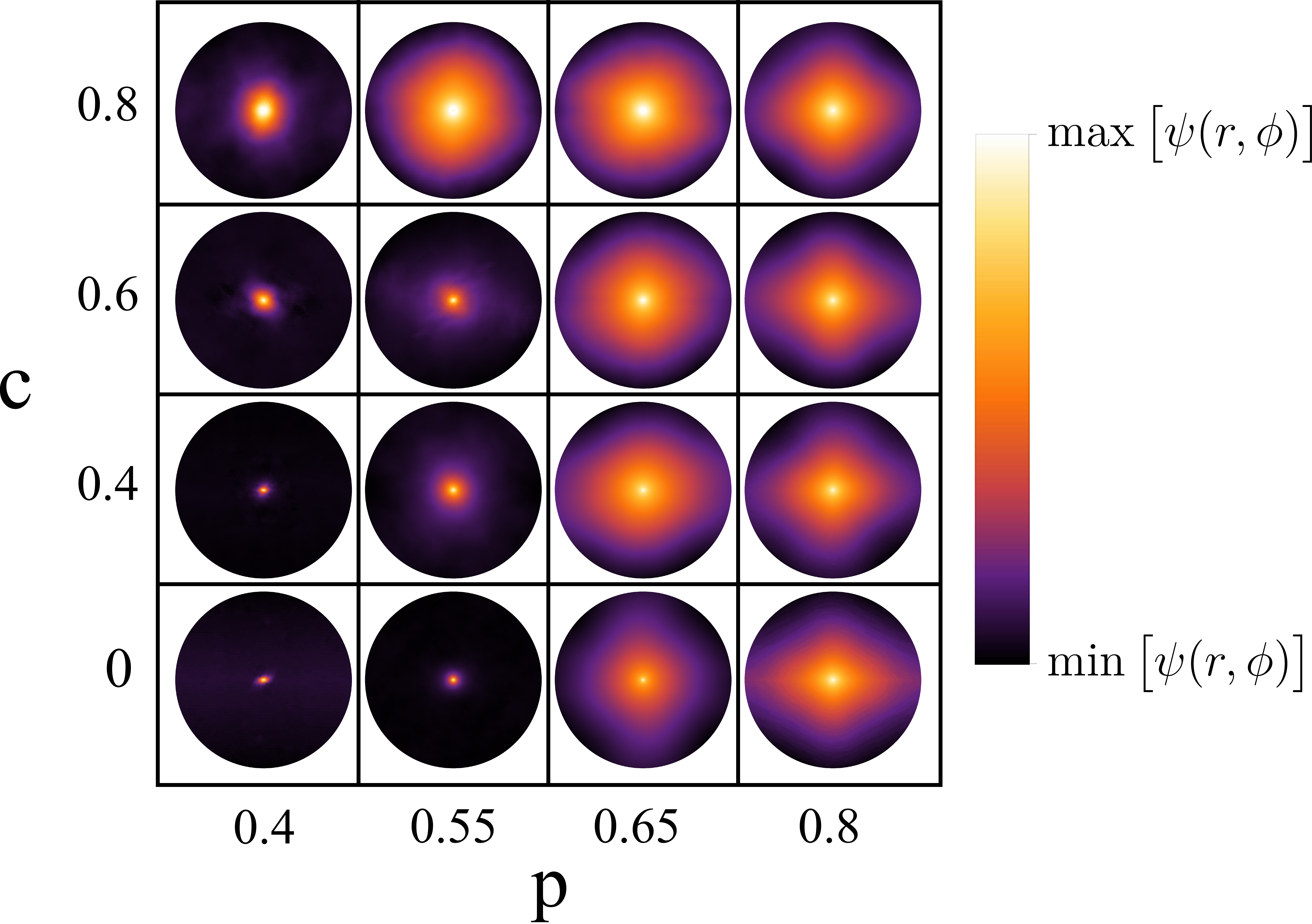}
    \caption{We show the correlation in non-affine parameter,
    as a function of the magnitude and orientation
    of separation. The color scale for each panel is normalized
    to the maximum value for a given combination of bond portion
    and correlation strength. The decay in non-affine 
    displacement coefficient becomes more gradual with growing
    bond portion for all correlation strength, but growth in
    decay length becomes markedly more rapid for highly
    correlated networks. We also note that, with growing
    bond portion, non-affine correlations exhibit increasing
    anisotropy, with decay in correlations becoming much more
    gradual along the direction of applied shear.}
    \label{fig:na_corr_polar}
\end{figure*}

We further look for evidence of orientational order in
non-affine correlations by averaging not over an annular
sector, but rather over all pairs of points whose relative
displacement has magnitude $r$, and makes an angle $\varphi$
with the positive $x$ axis. We define a measure of correlation
$\psi(r, \varphi)$:

\begin{equation}
    \psi(r, \phi) = \frac{\sum_{\vec{r}_1, \vec{r}_2} \vec{u}_{NA} (\vec{r}_1) \cdot \vec{u}_{NA}(\vec{r}_2) \delta_{r, |\vec{r}_1 - \vec{r}_2|} \delta_{\varphi, \theta_{1,2}}}{\left< \left| \vec{u}_{NA} \right|^2 \right> \sum_{\vec{r}_1, \vec{r}_2} \delta_{r, |\vec{r}_1 - \vec{r}_2|} \delta_{\varphi,\theta_{1,2}}},
\end{equation}
where summations are over all vertices, $\delta$ is the Kronecker delta,
and $\theta_{1,2}$ is the angle between the displacement 
$\vec{r}_2 - \vec{r}_2$ and the positive x axis. Results are shown in
Fig. ~\ref{fig:na_corr_polar}. The color scale for each combination
of correlation strength and bond portion is mapped to the range spanning
the minimum and maximum values of $\psi$ for that combination.

While for low bond portions, the correlation between
non-affine displacement is highly isotropic, correlations decay
far more gradually along the direction of applied strain at large
bond portion. Well above the rigidity percolation threshold,
networks deform in a nearly affine manner, as shown in Fig. ~\ref{fig:na_scaling} \textbf{a}. In this regime, the discrete
rotational symmetry of the Kagome network introduces elastic
anisotropy. This anisotropy is more pronounced for networks with less structural
correlation, as highly correlated networks have greater fluctuation in local
stiffness, a trait known to increase non-affinity ~\cite{DiDonna}.

\section*{Acknowledgments}
This research was supported in part by NSF grants DMR-1808026 and DMR-1807602. 

\providecommand{\noopsort}[1]{}\providecommand{\singleletter}[1]{#1}%


\begin{thebibliography}{33}%
\makeatletter
\providecommand \@ifxundefined [1]{%
 \@ifx{#1\undefined}
}%
\providecommand \@ifnum [1]{%
 \ifnum #1\expandafter \@firstoftwo
 \else \expandafter \@secondoftwo
 \fi
}%
\providecommand \@ifx [1]{%
 \ifx #1\expandafter \@firstoftwo
 \else \expandafter \@secondoftwo
 \fi
}%
\providecommand \natexlab [1]{#1}%
\providecommand \enquote  [1]{``#1''}%
\providecommand \bibnamefont  [1]{#1}%
\providecommand \bibfnamefont [1]{#1}%
\providecommand \citenamefont [1]{#1}%
\providecommand \href@noop [0]{\@secondoftwo}%
\providecommand \href [0]{\begingroup \@sanitize@url \@href}%
\providecommand \@href[1]{\@@startlink{#1}\@@href}%
\providecommand \@@href[1]{\endgroup#1\@@endlink}%
\providecommand \@sanitize@url [0]{\catcode `\\12\catcode `\$12\catcode
  `\&12\catcode `\#12\catcode `\^12\catcode `\_12\catcode `\%12\relax}%
\providecommand \@@startlink[1]{}%
\providecommand \@@endlink[0]{}%
\providecommand \url  [0]{\begingroup\@sanitize@url \@url }%
\providecommand \@url [1]{\endgroup\@href {#1}{\urlprefix }}%
\providecommand \urlprefix  [0]{URL }%
\providecommand \Eprint [0]{\href }%
\providecommand \doibase [0]{https://doi.org/}%
\providecommand \selectlanguage [0]{\@gobble}%
\providecommand \bibinfo  [0]{\@secondoftwo}%
\providecommand \bibfield  [0]{\@secondoftwo}%
\providecommand \translation [1]{[#1]}%
\providecommand \BibitemOpen [0]{}%
\providecommand \bibitemStop [0]{}%
\providecommand \bibitemNoStop [0]{.\EOS\space}%
\providecommand \EOS [0]{\spacefactor3000\relax}%
\providecommand \BibitemShut  [1]{\csname bibitem#1\endcsname}%
\let\auto@bib@innerbib\@empty
\bibitem [{\citenamefont {Pollard}\ and\ \citenamefont
  {Goldman}(2017)}]{Pollard}%
  \BibitemOpen
  \bibinfo {editor} {\bibfnamefont {T.~D.}\ \bibnamefont {Pollard}}\ and\
  \bibinfo {editor} {\bibfnamefont {R.~D.}\ \bibnamefont {Goldman}},\ eds.,\
  \href@noop {} {\emph {\bibinfo {title} {The Cytoskeleton}}}\ (\bibinfo
  {publisher} {Cold Spring Harbor Laboratory Press},\ \bibinfo {year}
  {2017})\BibitemShut {NoStop}%
\bibitem [{\citenamefont {Gardel}\ \emph
  {et~al.}(2004{\natexlab{a}})\citenamefont {Gardel}, \citenamefont {Shin},
  \citenamefont {MacKintosh}, \citenamefont {Mahadevan}, \citenamefont
  {Matsudaira},\ and\ \citenamefont {Weitz}}]{Gardel1}%
  \BibitemOpen
  \bibfield  {author} {\bibinfo {author} {\bibfnamefont {M.}~\bibnamefont
  {Gardel}}, \bibinfo {author} {\bibfnamefont {J.}~\bibnamefont {Shin}},
  \bibinfo {author} {\bibfnamefont {F.}~\bibnamefont {MacKintosh}}, \bibinfo
  {author} {\bibfnamefont {L.}~\bibnamefont {Mahadevan}}, \bibinfo {author}
  {\bibfnamefont {P.}~\bibnamefont {Matsudaira}},\ and\ \bibinfo {author}
  {\bibfnamefont {D.}~\bibnamefont {Weitz}},\ }\bibfield  {title} {\bibinfo
  {title} {Elastic behavior of cross-linked and bundled actin networks},\
  }\href@noop {} {\bibfield  {journal} {\bibinfo  {journal} {Science}\ }\textbf
  {\bibinfo {volume} {304}},\ \bibinfo {pages} {1301} (\bibinfo {year}
  {2004}{\natexlab{a}})}\BibitemShut {NoStop}%
\bibitem [{\citenamefont {Gardel}\ \emph
  {et~al.}(2004{\natexlab{b}})\citenamefont {Gardel}, \citenamefont {Shin},
  \citenamefont {MacKintosh}, \citenamefont {Mahadevan}, \citenamefont
  {Matsudaira},\ and\ \citenamefont {Weitz}}]{Gardel2}%
  \BibitemOpen
  \bibfield  {author} {\bibinfo {author} {\bibfnamefont {M.}~\bibnamefont
  {Gardel}}, \bibinfo {author} {\bibfnamefont {J.}~\bibnamefont {Shin}},
  \bibinfo {author} {\bibfnamefont {F.}~\bibnamefont {MacKintosh}}, \bibinfo
  {author} {\bibfnamefont {L.}~\bibnamefont {Mahadevan}}, \bibinfo {author}
  {\bibfnamefont {P.}~\bibnamefont {Matsudaira}},\ and\ \bibinfo {author}
  {\bibfnamefont {D.}~\bibnamefont {Weitz}},\ }\bibfield  {title} {\bibinfo
  {title} {Scaling of f-actin network rheology to probe single filament
  elasticity and dynamics},\ }\href@noop {} {\bibfield  {journal} {\bibinfo
  {journal} {Phys. Rev. Lett.}\ }\textbf {\bibinfo {volume} {93}},\ \bibinfo
  {pages} {188102} (\bibinfo {year} {2004}{\natexlab{b}})}\BibitemShut
  {NoStop}%
\bibitem [{\citenamefont {Murrell}\ \emph {et~al.}(2015)\citenamefont
  {Murrell}, \citenamefont {Oakes}, \citenamefont {Lenz},\ and\ \citenamefont
  {Gardel}}]{Murrell}%
  \BibitemOpen
  \bibfield  {author} {\bibinfo {author} {\bibfnamefont {M.}~\bibnamefont
  {Murrell}}, \bibinfo {author} {\bibfnamefont {P.~W.}\ \bibnamefont {Oakes}},
  \bibinfo {author} {\bibfnamefont {M.}~\bibnamefont {Lenz}},\ and\ \bibinfo
  {author} {\bibfnamefont {M.}~\bibnamefont {Gardel}},\ }\bibfield  {title}
  {\bibinfo {title} {Forcing cells into shape: the mechanics of actomyosin
  contractility},\ }\href@noop {} {\bibfield  {journal} {\bibinfo  {journal}
  {Nature Reviews Molecular Cell Biology}\ }\textbf {\bibinfo {volume} {16}},\
  \bibinfo {pages} {486} (\bibinfo {year} {2015})}\BibitemShut {NoStop}%
\bibitem [{\citenamefont {Lee}\ \emph {et~al.}(2021{\natexlab{a}})\citenamefont
  {Lee}, \citenamefont {Leech}, \citenamefont {Rust}, \citenamefont {Das},
  \citenamefont {McGorty}, \citenamefont {Ross},\ and\ \citenamefont
  {Robertson-Anderson}}]{Lee1}%
  \BibitemOpen
  \bibfield  {author} {\bibinfo {author} {\bibfnamefont {G.}~\bibnamefont
  {Lee}}, \bibinfo {author} {\bibfnamefont {G.}~\bibnamefont {Leech}}, \bibinfo
  {author} {\bibfnamefont {M.~J.}\ \bibnamefont {Rust}}, \bibinfo {author}
  {\bibfnamefont {M.}~\bibnamefont {Das}}, \bibinfo {author} {\bibfnamefont
  {R.~J.}\ \bibnamefont {McGorty}}, \bibinfo {author} {\bibfnamefont {J.~L.}\
  \bibnamefont {Ross}},\ and\ \bibinfo {author} {\bibfnamefont {R.~M.}\
  \bibnamefont {Robertson-Anderson}},\ }\bibfield  {title} {\bibinfo {title}
  {Myosin-driven actin-microtubule networks exhibit self-organized contractile
  dynamics},\ }\href@noop {} {\bibfield  {journal} {\bibinfo  {journal}
  {Science Advances}\ }\textbf {\bibinfo {volume} {7}},\ \bibinfo {pages}
  {4334} (\bibinfo {year} {2021}{\natexlab{a}})}\BibitemShut {NoStop}%
\bibitem [{\citenamefont {Lee}\ \emph {et~al.}(2021{\natexlab{b}})\citenamefont
  {Lee}, \citenamefont {Leech}, \citenamefont {Lwin}, \citenamefont {Michel},
  \citenamefont {Currie}, \citenamefont {Rust}, \citenamefont {Ross},
  \citenamefont {McGorty}, \citenamefont {Das},\ and\ \citenamefont
  {Robertson-Anderson}}]{Lee2}%
  \BibitemOpen
  \bibfield  {author} {\bibinfo {author} {\bibfnamefont {G.}~\bibnamefont
  {Lee}}, \bibinfo {author} {\bibfnamefont {G.}~\bibnamefont {Leech}}, \bibinfo
  {author} {\bibfnamefont {P.}~\bibnamefont {Lwin}}, \bibinfo {author}
  {\bibfnamefont {J.}~\bibnamefont {Michel}}, \bibinfo {author} {\bibfnamefont
  {C.}~\bibnamefont {Currie}}, \bibinfo {author} {\bibfnamefont {M.~J.}\
  \bibnamefont {Rust}}, \bibinfo {author} {\bibfnamefont {J.~L.}\ \bibnamefont
  {Ross}}, \bibinfo {author} {\bibfnamefont {R.~J.}\ \bibnamefont {McGorty}},
  \bibinfo {author} {\bibfnamefont {M.}~\bibnamefont {Das}},\ and\ \bibinfo
  {author} {\bibfnamefont {R.~M.}\ \bibnamefont {Robertson-Anderson}},\
  }\bibfield  {title} {\bibinfo {title} {Active cytoskeletal composites display
  emergent tunable contractility and restructuring},\ }\href
  {https://doi.org/10.1039/D1SM01083B} {\bibfield  {journal} {\bibinfo
  {journal} {Soft Matter}\ }\textbf {\bibinfo {volume} {17}},\ \bibinfo {pages}
  {10765} (\bibinfo {year} {2021}{\natexlab{b}})}\BibitemShut {NoStop}%
\bibitem [{\citenamefont {M{\"u}nster}\ \emph {et~al.}(2013)\citenamefont
  {M{\"u}nster}, \citenamefont {Jawerth}, \citenamefont {Leslie}, \citenamefont
  {Weitz}, \citenamefont {Fabry},\ and\ \citenamefont {Weitz}}]{Munster}%
  \BibitemOpen
  \bibfield  {author} {\bibinfo {author} {\bibfnamefont {S.}~\bibnamefont
  {M{\"u}nster}}, \bibinfo {author} {\bibfnamefont {L.~M.}\ \bibnamefont
  {Jawerth}}, \bibinfo {author} {\bibfnamefont {B.~A.}\ \bibnamefont {Leslie}},
  \bibinfo {author} {\bibfnamefont {J.~I.}\ \bibnamefont {Weitz}}, \bibinfo
  {author} {\bibfnamefont {B.}~\bibnamefont {Fabry}},\ and\ \bibinfo {author}
  {\bibfnamefont {D.~A.}\ \bibnamefont {Weitz}},\ }\bibfield  {title} {\bibinfo
  {title} {Strain history dependence of the nonlinear stress response of fibrin
  and collagen networks},\ }\href@noop {} {\bibfield  {journal} {\bibinfo
  {journal} {Proc. Nat. Acad. Sci. USA}\ }\textbf {\bibinfo {volume} {110}},\
  \bibinfo {pages} {12197} (\bibinfo {year} {2013})}\BibitemShut {NoStop}%
\bibitem [{\citenamefont {Silverberg}\ \emph {et~al.}(2014)\citenamefont
  {Silverberg}, \citenamefont {Barrett}, \citenamefont {Das}, \citenamefont
  {Petersen}, \citenamefont {Bonassar},\ and\ \citenamefont
  {Cohen}}]{silverberg_structure-function_2014}%
  \BibitemOpen
  \bibfield  {author} {\bibinfo {author} {\bibfnamefont {J.~L.}\ \bibnamefont
  {Silverberg}}, \bibinfo {author} {\bibfnamefont {A.~R.}\ \bibnamefont
  {Barrett}}, \bibinfo {author} {\bibfnamefont {M.}~\bibnamefont {Das}},
  \bibinfo {author} {\bibfnamefont {P.~B.}\ \bibnamefont {Petersen}}, \bibinfo
  {author} {\bibfnamefont {L.~J.}\ \bibnamefont {Bonassar}},\ and\ \bibinfo
  {author} {\bibfnamefont {I.}~\bibnamefont {Cohen}},\ }\bibfield  {title}
  {\bibinfo {title} {Structure-function relations and rigidity percolation in
  the shear properties of articular cartilage},\ }\href
  {https://doi.org/10.1016/j.bpj.2014.08.011} {\bibfield  {journal} {\bibinfo
  {journal} {Biophysical Journal}\ }\textbf {\bibinfo {volume} {107}},\
  \bibinfo {pages} {1721} (\bibinfo {year} {2014})}\BibitemShut {NoStop}%
\bibitem [{\citenamefont {Jansen}\ \emph {et~al.}(2018)\citenamefont {Jansen},
  \citenamefont {Licup}, \citenamefont {Sharma}, \citenamefont {Rens},
  \citenamefont {MacKintosh},\ and\ \citenamefont {Koenderink}}]{KAJansen}%
  \BibitemOpen
  \bibfield  {author} {\bibinfo {author} {\bibfnamefont {K.~A.}\ \bibnamefont
  {Jansen}}, \bibinfo {author} {\bibfnamefont {A.~J.}\ \bibnamefont {Licup}},
  \bibinfo {author} {\bibfnamefont {A.}~\bibnamefont {Sharma}}, \bibinfo
  {author} {\bibfnamefont {R.}~\bibnamefont {Rens}}, \bibinfo {author}
  {\bibfnamefont {F.~C.}\ \bibnamefont {MacKintosh}},\ and\ \bibinfo {author}
  {\bibfnamefont {G.~H.}\ \bibnamefont {Koenderink}},\ }\bibfield  {title}
  {\bibinfo {title} {The role of network architecture in collagen mechanics},\
  }\href@noop {} {\bibfield  {journal} {\bibinfo  {journal} {Biophysical
  Journal}\ }\textbf {\bibinfo {volume} {114}},\ \bibinfo {pages} {2665}
  (\bibinfo {year} {2018})}\BibitemShut {NoStop}%
\bibitem [{\citenamefont {Burla}\ \emph {et~al.}(2020)\citenamefont {Burla},
  \citenamefont {Dussi}, \citenamefont {Martinez-Torres}, \citenamefont
  {Tauber}, \citenamefont {van~der Gucht},\ and\ \citenamefont
  {Koenderink}}]{Burla1}%
  \BibitemOpen
  \bibfield  {author} {\bibinfo {author} {\bibfnamefont {F.}~\bibnamefont
  {Burla}}, \bibinfo {author} {\bibfnamefont {C.}~\bibnamefont {Dussi}},
  \bibinfo {author} {\bibfnamefont {J.}~\bibnamefont {Martinez-Torres}},
  \bibinfo {author} {\bibfnamefont {J.}~\bibnamefont {Tauber}}, \bibinfo
  {author} {\bibfnamefont {J.}~\bibnamefont {van~der Gucht}},\ and\ \bibinfo
  {author} {\bibfnamefont {G.~H.}\ \bibnamefont {Koenderink}},\ }\bibfield
  {title} {\bibinfo {title} {Connectivity and plasticity determine collagen
  network fracture},\ }\href@noop {} {\bibfield  {journal} {\bibinfo  {journal}
  {Proc. Nat. Acad. Sci. USA}\ }\textbf {\bibinfo {volume} {117}},\ \bibinfo
  {pages} {8326} (\bibinfo {year} {2020})}\BibitemShut {NoStop}%
\bibitem [{\citenamefont {Wyse~Jackson}\ \emph {et~al.}(2022)\citenamefont
  {Wyse~Jackson}, \citenamefont {Michel}, \citenamefont {Lwin}, \citenamefont
  {Fortier}, \citenamefont {Das}, \citenamefont {Bonassar},\ and\ \citenamefont
  {Cohen}}]{Wyse_Jackson}%
  \BibitemOpen
  \bibfield  {author} {\bibinfo {author} {\bibfnamefont {T.}~\bibnamefont
  {Wyse~Jackson}}, \bibinfo {author} {\bibfnamefont {J.~A.}\ \bibnamefont
  {Michel}}, \bibinfo {author} {\bibfnamefont {P.}~\bibnamefont {Lwin}},
  \bibinfo {author} {\bibfnamefont {L.~A.}\ \bibnamefont {Fortier}}, \bibinfo
  {author} {\bibfnamefont {M.}~\bibnamefont {Das}}, \bibinfo {author}
  {\bibfnamefont {L.~J.}\ \bibnamefont {Bonassar}},\ and\ \bibinfo {author}
  {\bibfnamefont {I.}~\bibnamefont {Cohen}},\ }\bibfield  {title} {\bibinfo
  {title} {Structural origins of cartilage shear mechanics},\ }\href@noop {}
  {\bibfield  {journal} {\bibinfo  {journal} {Science Advances}\ }\textbf
  {\bibinfo {volume} {8}},\ \bibinfo {pages} {2805} (\bibinfo {year}
  {2022})}\BibitemShut {NoStop}%
\bibitem [{\citenamefont {Petridou}\ \emph {et~al.}(2021)\citenamefont
  {Petridou}, \citenamefont {Corominas-Murta}, \citenamefont {Heisenberg},\
  and\ \citenamefont {Hannezo}}]{Heisenberg}%
  \BibitemOpen
  \bibfield  {author} {\bibinfo {author} {\bibfnamefont {N.~I.}\ \bibnamefont
  {Petridou}}, \bibinfo {author} {\bibfnamefont {B.}~\bibnamefont
  {Corominas-Murta}}, \bibinfo {author} {\bibfnamefont {C.}~\bibnamefont
  {Heisenberg}},\ and\ \bibinfo {author} {\bibfnamefont {E.}~\bibnamefont
  {Hannezo}},\ }\bibfield  {title} {\bibinfo {title} {Rigidity percolation
  uncovers a structural basis for embryonic tissue phase transitions},\
  }\href@noop {} {\bibfield  {journal} {\bibinfo  {journal} {Cell}\ }\textbf
  {\bibinfo {volume} {184}},\ \bibinfo {pages} {1914} (\bibinfo {year}
  {2021})}\BibitemShut {NoStop}%
\bibitem [{\citenamefont {DiDomenico}\ \emph {et~al.}(2018)\citenamefont
  {DiDomenico}, \citenamefont {Lintz},\ and\ \citenamefont
  {Bonassar}}]{DiDomenico}%
  \BibitemOpen
  \bibfield  {author} {\bibinfo {author} {\bibfnamefont {C.~D.}\ \bibnamefont
  {DiDomenico}}, \bibinfo {author} {\bibfnamefont {M.}~\bibnamefont {Lintz}},\
  and\ \bibinfo {author} {\bibfnamefont {L.~J.}\ \bibnamefont {Bonassar}},\
  }\bibfield  {title} {\bibinfo {title} {Molecular transport in articular
  cartilage - what have we learned from the last 50 years?},\ }\href@noop {}
  {\bibfield  {journal} {\bibinfo  {journal} {Nature Reviews Rheumatology}\
  }\textbf {\bibinfo {volume} {14}},\ \bibinfo {pages} {393} (\bibinfo {year}
  {2018})}\BibitemShut {NoStop}%
\bibitem [{\citenamefont {Rhee}\ \emph {et~al.}(2016)\citenamefont {Rhee},
  \citenamefont {Puetzer}, \citenamefont {Mason}, \citenamefont
  {Reinhart-King},\ and\ \citenamefont {Bonassar}}]{Rhee}%
  \BibitemOpen
  \bibfield  {author} {\bibinfo {author} {\bibfnamefont {S.}~\bibnamefont
  {Rhee}}, \bibinfo {author} {\bibfnamefont {J.~L.}\ \bibnamefont {Puetzer}},
  \bibinfo {author} {\bibfnamefont {B.~N.}\ \bibnamefont {Mason}}, \bibinfo
  {author} {\bibfnamefont {C.~A.}\ \bibnamefont {Reinhart-King}},\ and\
  \bibinfo {author} {\bibfnamefont {L.~J.}\ \bibnamefont {Bonassar}},\
  }\bibfield  {title} {\bibinfo {title} {3d bioprinting of spatially
  heterogeneous collagen constructs for cartilage tissue engineering},\
  }\href@noop {} {\bibfield  {journal} {\bibinfo  {journal} {ACS Biomater. Sci.
  Eng.}\ }\textbf {\bibinfo {volume} {2}},\ \bibinfo {pages} {1800} (\bibinfo
  {year} {2016})}\BibitemShut {NoStop}%
\bibitem [{\citenamefont {Lalitha~Sridhar}\ \emph {et~al.}(2017)\citenamefont
  {Lalitha~Sridhar}, \citenamefont {Schneider}, \citenamefont {Chu},
  \citenamefont {de~Roucy}, \citenamefont {Bryant},\ and\ \citenamefont
  {Vernerey}}]{Sridhar}%
  \BibitemOpen
  \bibfield  {author} {\bibinfo {author} {\bibfnamefont {S.}~\bibnamefont
  {Lalitha~Sridhar}}, \bibinfo {author} {\bibfnamefont {M.~C.}\ \bibnamefont
  {Schneider}}, \bibinfo {author} {\bibfnamefont {S.}~\bibnamefont {Chu}},
  \bibinfo {author} {\bibfnamefont {G.}~\bibnamefont {de~Roucy}}, \bibinfo
  {author} {\bibfnamefont {S.~J.}\ \bibnamefont {Bryant}},\ and\ \bibinfo
  {author} {\bibfnamefont {F.~J.}\ \bibnamefont {Vernerey}},\ }\bibfield
  {title} {\bibinfo {title} {Heterogeneity is key to hydrogel-based cartilage
  tissue regeneration},\ }\href {https://doi.org/10.1039/C7SM00423K} {\bibfield
   {journal} {\bibinfo  {journal} {Soft Matter}\ }\textbf {\bibinfo {volume}
  {13}},\ \bibinfo {pages} {4841} (\bibinfo {year} {2017})}\BibitemShut
  {NoStop}%
\bibitem [{\citenamefont {Head}\ \emph {et~al.}(2003)\citenamefont {Head},
  \citenamefont {Levine},\ and\ \citenamefont {MacKintsoh}}]{Head}%
  \BibitemOpen
  \bibfield  {author} {\bibinfo {author} {\bibfnamefont {D.~A.}\ \bibnamefont
  {Head}}, \bibinfo {author} {\bibfnamefont {A.~J.}\ \bibnamefont {Levine}},\
  and\ \bibinfo {author} {\bibfnamefont {F.~C.}\ \bibnamefont {MacKintsoh}},\
  }\bibfield  {title} {\bibinfo {title} {Distinct regimes of elastic response
  and deformation modes of cross-linked cytoskeletal and semiflexible polymer
  networks},\ }\href@noop {} {\bibfield  {journal} {\bibinfo  {journal} {Phys.
  Rev. E}\ }\textbf {\bibinfo {volume} {68}},\ \bibinfo {pages} {061907}
  (\bibinfo {year} {2003})}\BibitemShut {NoStop}%
\bibitem [{\citenamefont {Wilhelm}\ and\ \citenamefont {Frey}(2003)}]{Wilhelm}%
  \BibitemOpen
  \bibfield  {author} {\bibinfo {author} {\bibfnamefont {J.}~\bibnamefont
  {Wilhelm}}\ and\ \bibinfo {author} {\bibfnamefont {E.}~\bibnamefont {Frey}},\
  }\bibfield  {title} {\bibinfo {title} {Elasticity of stiff polymer
  networks},\ }\href@noop {} {\bibfield  {journal} {\bibinfo  {journal} {Phys.
  Rev. Lett.}\ }\textbf {\bibinfo {volume} {91}},\ \bibinfo {pages} {108103}
  (\bibinfo {year} {2003})}\BibitemShut {NoStop}%
\bibitem [{\citenamefont {Heussinger}\ and\ \citenamefont
  {Frey}(2006)}]{Heussinger}%
  \BibitemOpen
  \bibfield  {author} {\bibinfo {author} {\bibfnamefont {C.}~\bibnamefont
  {Heussinger}}\ and\ \bibinfo {author} {\bibfnamefont {E.}~\bibnamefont
  {Frey}},\ }\bibfield  {title} {\bibinfo {title} {Floppy modes and nonaffine
  deformations in random fiber networks},\ }\href@noop {} {\bibfield  {journal}
  {\bibinfo  {journal} {Phys. Rev. Lett.}\ }\textbf {\bibinfo {volume} {97}},\
  \bibinfo {pages} {105501} (\bibinfo {year} {2006})}\BibitemShut {NoStop}%
\bibitem [{\citenamefont {Das}\ \emph {et~al.}(2007)\citenamefont {Das},
  \citenamefont {MacKintosh},\ and\ \citenamefont {Levine}}]{Das1}%
  \BibitemOpen
  \bibfield  {author} {\bibinfo {author} {\bibfnamefont {M.}~\bibnamefont
  {Das}}, \bibinfo {author} {\bibfnamefont {F.~C.}\ \bibnamefont
  {MacKintosh}},\ and\ \bibinfo {author} {\bibfnamefont {A.~J.}\ \bibnamefont
  {Levine}},\ }\bibfield  {title} {\bibinfo {title} {Effective medium theory of
  semiflexible filamentous networks},\ }\href@noop {} {\bibfield  {journal}
  {\bibinfo  {journal} {Physical Review Letters}\ }\textbf {\bibinfo {volume}
  {99}},\ \bibinfo {pages} {038101} (\bibinfo {year} {2007})}\BibitemShut
  {NoStop}%
\bibitem [{\citenamefont {Huisman}\ \emph {et~al.}(2007)\citenamefont
  {Huisman}, \citenamefont {van Dillen}, \citenamefont {Onck},\ and\
  \citenamefont {Van~der Giessen}}]{Huisman}%
  \BibitemOpen
  \bibfield  {author} {\bibinfo {author} {\bibfnamefont {E.~M.}\ \bibnamefont
  {Huisman}}, \bibinfo {author} {\bibfnamefont {T.}~\bibnamefont {van Dillen}},
  \bibinfo {author} {\bibfnamefont {P.~R.}\ \bibnamefont {Onck}},\ and\
  \bibinfo {author} {\bibfnamefont {E.}~\bibnamefont {Van~der Giessen}},\
  }\bibfield  {title} {\bibinfo {title} {Three-dimensional cross-linked f-actin
  networks: Relation between network architecture and mechanical behavior},\
  }\href@noop {} {\bibfield  {journal} {\bibinfo  {journal} {Physical Review
  Letters}\ }\textbf {\bibinfo {volume} {99}},\ \bibinfo {pages} {208103}
  (\bibinfo {year} {2007})}\BibitemShut {NoStop}%
\bibitem [{\citenamefont {Broedersz}\ \emph {et~al.}(2011)\citenamefont
  {Broedersz}, \citenamefont {Mao}, \citenamefont {Lubensky},\ and\
  \citenamefont {MacKintosh}}]{Broedersz1}%
  \BibitemOpen
  \bibfield  {author} {\bibinfo {author} {\bibfnamefont {C.}~\bibnamefont
  {Broedersz}}, \bibinfo {author} {\bibfnamefont {X.}~\bibnamefont {Mao}},
  \bibinfo {author} {\bibfnamefont {T.}~\bibnamefont {Lubensky}},\ and\
  \bibinfo {author} {\bibfnamefont {F.~C.}\ \bibnamefont {MacKintosh}},\
  }\bibfield  {title} {\bibinfo {title} {Criticality and isostaticity in fibre
  networks},\ }\href@noop {} {\bibfield  {journal} {\bibinfo  {journal} {Nature
  Physics}\ }\textbf {\bibinfo {volume} {7}},\ \bibinfo {pages} {983} (\bibinfo
  {year} {2011})}\BibitemShut {NoStop}%
\bibitem [{\citenamefont {Das}\ \emph {et~al.}(2012)\citenamefont {Das},
  \citenamefont {Quint},\ and\ \citenamefont {Schwarz}}]{Das2}%
  \BibitemOpen
  \bibfield  {author} {\bibinfo {author} {\bibfnamefont {M.}~\bibnamefont
  {Das}}, \bibinfo {author} {\bibfnamefont {D.~A.}\ \bibnamefont {Quint}},\
  and\ \bibinfo {author} {\bibfnamefont {J.~M.}\ \bibnamefont {Schwarz}},\
  }\bibfield  {title} {\bibinfo {title} {Redundancy and cooperativity in the
  mechanics of compositely crosslinked filamentous networks},\ }\href@noop {}
  {\bibfield  {journal} {\bibinfo  {journal} {PLoS ONE}\ }\textbf {\bibinfo
  {volume} {7}},\ \bibinfo {pages} {35939} (\bibinfo {year}
  {2012})}\BibitemShut {NoStop}%
\bibitem [{\citenamefont {Picu}(2011)}]{Picu}%
  \BibitemOpen
  \bibfield  {author} {\bibinfo {author} {\bibfnamefont {R.~C.}\ \bibnamefont
  {Picu}},\ }\bibfield  {title} {\bibinfo {title} {Mechanics of random fiber
  networks—a review},\ }\href@noop {} {\bibfield  {journal} {\bibinfo
  {journal} {Soft Matter}\ }\textbf {\bibinfo {volume} {7}},\ \bibinfo {pages}
  {6768} (\bibinfo {year} {2011})}\BibitemShut {NoStop}%
\bibitem [{\citenamefont {Broedersz}\ and\ \citenamefont
  {MacKintosh}(2014)}]{Broedersz2}%
  \BibitemOpen
  \bibfield  {author} {\bibinfo {author} {\bibfnamefont {C.~P.}\ \bibnamefont
  {Broedersz}}\ and\ \bibinfo {author} {\bibfnamefont {F.~C.}\ \bibnamefont
  {MacKintosh}},\ }\bibfield  {title} {\bibinfo {title} {Modeling semiflexible
  polymer networks},\ }\href@noop {} {\bibfield  {journal} {\bibinfo  {journal}
  {Rev. Mod. Phys.}\ }\textbf {\bibinfo {volume} {86}},\ \bibinfo {pages} {995}
  (\bibinfo {year} {2014})}\BibitemShut {NoStop}%
\bibitem [{\citenamefont {Feng}\ \emph {et~al.}(1985)\citenamefont {Feng},
  \citenamefont {Thorpe},\ and\ \citenamefont {Garboczi}}]{Feng}%
  \BibitemOpen
  \bibfield  {author} {\bibinfo {author} {\bibfnamefont {S.}~\bibnamefont
  {Feng}}, \bibinfo {author} {\bibfnamefont {M.~F.}\ \bibnamefont {Thorpe}},\
  and\ \bibinfo {author} {\bibfnamefont {E.}~\bibnamefont {Garboczi}},\
  }\bibfield  {title} {\bibinfo {title} {Effective-medium theory of percolation
  on central-force elastic networks},\ }\href@noop {} {\bibfield  {journal}
  {\bibinfo  {journal} {Phys. Rev. B}\ }\textbf {\bibinfo {volume} {31}},\
  \bibinfo {pages} {276} (\bibinfo {year} {1985})}\BibitemShut {NoStop}%
\bibitem [{\citenamefont {Storm}\ \emph {et~al.}(2005)\citenamefont {Storm},
  \citenamefont {Pastore}, \citenamefont {MacKintosh}, \citenamefont
  {Lubensky},\ and\ \citenamefont {Janmey}}]{Storm}%
  \BibitemOpen
  \bibfield  {author} {\bibinfo {author} {\bibfnamefont {C.}~\bibnamefont
  {Storm}}, \bibinfo {author} {\bibfnamefont {J.~J.}\ \bibnamefont {Pastore}},
  \bibinfo {author} {\bibfnamefont {F.~C.}\ \bibnamefont {MacKintosh}},
  \bibinfo {author} {\bibfnamefont {T.~C.}\ \bibnamefont {Lubensky}},\ and\
  \bibinfo {author} {\bibfnamefont {P.~A.}\ \bibnamefont {Janmey}},\ }\bibfield
   {title} {\bibinfo {title} {Nonlinear elasticity in biological gels},\
  }\href@noop {} {\bibfield  {journal} {\bibinfo  {journal} {Nature}\ }\textbf
  {\bibinfo {volume} {194}},\ \bibinfo {pages} {191} (\bibinfo {year}
  {2005})}\BibitemShut {NoStop}%
\bibitem [{\citenamefont {Zhang}\ \emph {et~al.}(2019)\citenamefont {Zhang},
  \citenamefont {Zhang}, \citenamefont {Bouzid}, \citenamefont {Rocklin},
  \citenamefont {Del~Gado},\ and\ \citenamefont {Mao}}]{zhang_corr}%
  \BibitemOpen
  \bibfield  {author} {\bibinfo {author} {\bibfnamefont {S.}~\bibnamefont
  {Zhang}}, \bibinfo {author} {\bibfnamefont {L.}~\bibnamefont {Zhang}},
  \bibinfo {author} {\bibfnamefont {M.}~\bibnamefont {Bouzid}}, \bibinfo
  {author} {\bibfnamefont {D.~Z.}\ \bibnamefont {Rocklin}}, \bibinfo {author}
  {\bibfnamefont {E.}~\bibnamefont {Del~Gado}},\ and\ \bibinfo {author}
  {\bibfnamefont {X.}~\bibnamefont {Mao}},\ }\bibfield  {title} {\bibinfo
  {title} {Correlated rigidity percolation and colloidal gels},\ }\href@noop {}
  {\bibfield  {journal} {\bibinfo  {journal} {Phys. Rev. Lett.}\ }\textbf
  {\bibinfo {volume} {123}},\ \bibinfo {pages} {058001} (\bibinfo {year}
  {2019})}\BibitemShut {NoStop}%
\bibitem [{\citenamefont {DiDonna}\ and\ \citenamefont
  {Lubensky}(2005)}]{DiDonna}%
  \BibitemOpen
  \bibfield  {author} {\bibinfo {author} {\bibfnamefont {B.~A.}\ \bibnamefont
  {DiDonna}}\ and\ \bibinfo {author} {\bibfnamefont {T.~C.}\ \bibnamefont
  {Lubensky}},\ }\bibfield  {title} {\bibinfo {title} {Nonaffine correlations
  in random elastic media},\ }\href@noop {} {\bibfield  {journal} {\bibinfo
  {journal} {Phys. Rev. E}\ }\textbf {\bibinfo {volume} {72}},\ \bibinfo
  {pages} {066619} (\bibinfo {year} {2005})}\BibitemShut {NoStop}%
\bibitem [{\citenamefont {Zhang}\ \emph {et~al.}(2014)\citenamefont {Zhang},
  \citenamefont {Schwarz},\ and\ \citenamefont {Das}}]{Zhang_Aniso}%
  \BibitemOpen
  \bibfield  {author} {\bibinfo {author} {\bibfnamefont {T.}~\bibnamefont
  {Zhang}}, \bibinfo {author} {\bibfnamefont {J.~M.}\ \bibnamefont {Schwarz}},\
  and\ \bibinfo {author} {\bibfnamefont {M.}~\bibnamefont {Das}},\ }\bibfield
  {title} {\bibinfo {title} {Mechanics of anisotropic spring networks},\
  }\href@noop {} {\bibfield  {journal} {\bibinfo  {journal} {Phys. Rev. E}\
  }\textbf {\bibinfo {volume} {90}},\ \bibinfo {pages} {062139} (\bibinfo
  {year} {2014})}\BibitemShut {NoStop}%
\bibitem [{\citenamefont {Moore}(1920)}]{Moore}%
  \BibitemOpen
  \bibfield  {author} {\bibinfo {author} {\bibfnamefont {E.~H.}\ \bibnamefont
  {Moore}},\ }\bibfield  {title} {\bibinfo {title} {On the reciprocal of the
  general algebraic matrix},\ }\href@noop {} {\bibfield  {journal} {\bibinfo
  {journal} {Bulletin of the American Mathematical Society}\ }\textbf {\bibinfo
  {volume} {26}},\ \bibinfo {pages} {394–395} (\bibinfo {year}
  {1920})}\BibitemShut {NoStop}%
\bibitem [{\citenamefont {Bjerhammar}(1951)}]{Bjerhammar}%
  \BibitemOpen
  \bibfield  {author} {\bibinfo {author} {\bibfnamefont {A.}~\bibnamefont
  {Bjerhammar}},\ }\bibfield  {title} {\bibinfo {title} {Application of
  calculus of matrices to method of least squares: with special references to
  geodetic calculations},\ }\href@noop {} {\bibfield  {journal} {\bibinfo
  {journal} {Trans. Roy. Inst. Tech. Stockholm.}\ }\textbf {\bibinfo {volume}
  {49}} (\bibinfo {year} {1951})}\BibitemShut {NoStop}%
\bibitem [{\citenamefont {Penose}(1955)}]{Penrose}%
  \BibitemOpen
  \bibfield  {author} {\bibinfo {author} {\bibfnamefont {R.}~\bibnamefont
  {Penose}},\ }\bibfield  {title} {\bibinfo {title} {A generalized inverse for
  matrices},\ }\href@noop {} {\bibfield  {journal} {\bibinfo  {journal}
  {Mathematical Proceedings of the Cambridge Philosophical Society}\ }\textbf
  {\bibinfo {volume} {51}},\ \bibinfo {pages} {406} (\bibinfo {year}
  {1955})}\BibitemShut {NoStop}%
\bibitem [{\citenamefont {Davis}(2011)}]{Davis}%
  \BibitemOpen
  \bibfield  {author} {\bibinfo {author} {\bibfnamefont {T.~A.}\ \bibnamefont
  {Davis}},\ }\bibfield  {title} {\bibinfo {title} {Suitesparse{QR}:
  Multifrontal multithreaded rank-revealing sparse {QR} factorization},\
  }\href@noop {} {\bibfield  {journal} {\bibinfo  {journal} {ACM Transactions
  on Mathematical Software}\ }\textbf {\bibinfo {volume} {38}},\ \bibinfo
  {pages} {8:1 } (\bibinfo {year} {2011})}\BibitemShut {NoStop}%
\end{thebibliography}
\end{document}